\documentclass[
]{eptcs}
\providecommand{\event}{Linearity'16}

\usepackage[T1]{fontenc}
\usepackage{lmodern}
\usepackage{MnSymbol} 
\usepackage{cancel}
\usepackage{wrapfig}
\usepackage{subcaption}
\usepackage{adjustbox}

\usepackage{hyperref}

\usepackage[
    backend=bibtex,
    doi=true,
]{biblatex}
\usepackage{fixme}
\fxsetup {
  status=draft,
  layout=inline,
  theme=color
}

\usepackage{tikz}
\usetikzlibrary{shapes, positioning, arrows, arrows.meta}

\usepackage{listings}

\definecolor{cc0-green}{RGB}{0, 132, 61}
\definecolor{cc0-yellow}{RGB}{180, 125, 0}
\definecolor{cc0-red}{RGB}{180, 0, 0}
\definecolor{cc0-blue}{RGB}{0, 180, 0}
\definecolor{cc0-gray}{rgb}{0.5,0.5,0.5}

\lstdefinelanguage{C0} {
    language=c,
    commentstyle=\itshape\color{cc0-gray},
    keywords={},
    otherkeywords={!,?},
    keywordstyle=\bfseries, 
    keywordstyle=[2]\itshape , 
    morekeywords=[2]{int, string, choice, bool, void, struct, enum},
    keywordstyle=[3]\bfseries, 
    morekeywords=[3]{if, else, for, while, typedef, switch, case, true, false,
      return, break},
    keywordstyle=[4]\underline, 
    morekeywords=[4]{send, recv, wait, close},
}

\lstdefinelanguage{Go}%
{morekeywords=[1]{package,import,func,type,struct,return,defer,panic,%
    recover,select,var,const,iota,},%
  morekeywords=[2]{string,uint,uint8,uint16,uint32,uint64,int,int8,int16,%
    int32,int64,bool,float32,float64,complex64,complex128,byte,rune,uintptr,%
    error,interface},%
  morekeywords=[3]{map,slice,make,new,nil,len,cap,copy,close,true,false,%
    delete,append,real,imag,complex,chan,},%
  morekeywords=[4]{for,break,continue,range,goto,switch,case,fallthrough,if,%
    else,default,},%
  morekeywords=[5]{Println,Printf,Error,},%
  sensitive=true,%
  morecomment=[l]{//},%
  morecomment=[s]{/*}{*/},%
  morestring=[b]',%
  morestring=[b]",%
  morestring=[s]{`}{`},%
}

\title{Design and Implementation of Concurrent C0
  \thanks{An extended version of this paper can be found at
    \url{http://maxwillsey.com/papers/cc0-thesis.pdf}}}
\author{
  Max Willsey
  \institute{Carnegie Mellon University}
  \email{\href{mailto:mwillsey@cmu.edu}{mwillsey@cmu.edu}}
  \and
  Rokhini Prabhu
  \institute{Carnegie Mellon University}
  \email{\href{mailto:rokhini@gmail.com}{rokhini@gmail.com}}
  \and
  Frank Pfenning
  \institute{Carnegie Mellon University}
  \thanks{This work is partially funded by the FCT (Portuguese Foundation
  for Science and Technology) through the Carnegie Mellon
  Portugal Program.}
  \email{\href{mailto:fp@cmu.edu}{fp@cmu.edu}}
}

\date{Last updated: \today}

\def\titlerunning{Design and Implementation of Concurrent C0}
\def\authorrunning{M. Willsey, R. Prabhu \& F. Pfenning}

\begin{document}

\maketitle

\begin{abstract}

We describe Concurrent C0, a type-safe C-like language with contracts and
session-typed communication over channels. Concurrent C0 supports an operation
called forwarding which allows channels to be combined in a well-defined way.
The language's type system enables elegant expression of session types and
message-passing concurrent programs. We provide a Go-based implementation with
language based optimizations that outperforms traditional message passing
techniques.

\end{abstract}




\newcommand\intchoice\&
\newcommand\extchoice\oplus

\section{Introduction}

Message passing is an approach to concurrent programming where processes
do not operate directly on a shared state but instead communicate by passing
messages over channels. Many modern languages like Go, Rust, and Haskell provide
concurrent processes and channels to facilitate safe concurrent programming
through message passing, eliminating the need for locks in common communication
patterns. Most message passing systems implement \emph{asynchronous}
communication, where the channel contains a buffer so the sender can store the
message and proceed without waiting for it to be received.

However, conventional channels do not easily enable safe bidirectional
communication. Send\-ers must somehow ensure that they do not receive messages
that they just sent over the channel. Furthermore, complex protocols involve
multiple types of data, so statically typed channels must be created with the
sum of those types. The receiver must check the actual type of the value,
typically producing errors if the type is unexpected, essentially degenerating
to dynamic typing.

We propose the Concurrent C0 language as a tool to enable safer, more efficient
concurrent programming. Like other modern languages, it provides concurrent
processes that communicate over channels. It uses session typing to guarantee
the safety of communication and also to alleviate the burden of manually
synchronizing bidirectional communication \cite{Caires10concur, Honda93concur,
Toninho15phd}. Furthermore, Concurrent C0 offers a concise syntax to express
session typed protocols and programs adhering to them. The forwarding operation
creates ways to write programs not possible in other languages with message
passing. These language features provide additional safety and also enable an
optimized implementation.


\section{Concurrent C0}
\label{sec:lang}

Concurrent C0 is based on C0, an imperative programming language closely
resembling C designed for use in an introductory programming course. C0 intends
to have fully specified semantics to avoid the confusion that comes along with
C's undefined behavior \cite{Arnold10ms}. C0 provides memory safety by
disallowing pointer arithmetic and casting; all pointers come from the built-in
\texttt{alloc} and all arrays from the built-in \texttt{alloc\_array}, and they
are not interchangeable as in C. The C0 runtime \texttt{NULL}-checks pointer
accesses and bound-checks array accesses. C0 is garbage collected \cite{Boehm},
eliminating the need to explicitly free memory. C0 also supports optional
dynamically checked contracts of the familiar forms \texttt{@requires},
\texttt{@ensures}, and \texttt{@assert}. Students use these contracts to learn
how to reason about their code; in particular, the special
\texttt{@loop\_invariant} form allows students to reason about their loops. For
more on C0, see \cite{C0site}.

Concurrent C0 (CC0) is an extension of C0,
providing safety for the sequential aspects of programs written in CC0. The
session-typed concurrent extension is delimited from the sequential language,
and thus this paper's contributions could be be readily applied to any language
with a similar session-typed linear semantics.

\subsection{Concurrency}

\begin{wrapfigure}{R}{0.4\textwidth}
  \scriptsize
\begin{lstlisting}[xleftmargin=0.5cm,aboveskip=-0.8cm]
<!int;> $c fib(int n) {
    if (n == 0) {
	send($c, 0); close($c);
    } else if (n == 1) {
	send($c, 1); close($c);
    } else {
	<!int;> $c1 = fib(n-1);
	<!int;> $c2 = fib(n-2);
	int f1 = recv($c1); wait($c1);
	int f2 = recv($c2); wait($c2);
	send($c, f1+f2);    close($c);
    }
}
int main() {
    <!int;> $c = fib(10);
    int f = recv($c); wait($c);
    assert(f == 55);
    return 0;
}
\end{lstlisting}
  \caption{Naive concurrent Fibonacci.}
  \label{fig:fib.c1}
\end{wrapfigure}

Concurrent C0 extends C0 with the ability to create concurrent processes and
channels to communicate between them. In CC0, a \emph{process}\footnote{
  In the current implementations, \emph{processes} are units of
  execution within a single operating system process, and channels are
  implemented in shared memory. The features of Concurrent C0 generalize to any
  communicating processes, but this paper focuses on a shared memory
  implementation. For other applications, see \autoref{sec:distributed}.}
is a unit of concurrent execution, and \emph{channels} are effectively\footnote{
  For some session types, a bounded buffer can behave the
  same as an unbounded one. See \autoref{sec:type-width}.}
unbounded message buffers that allow the processes on either end to communicate
asynchronously.

Consider line 1 of \autoref{fig:fib.c1}: \lstinline+fib+ is a \emph{spawning
function} that creates and returns the channel \lstinline+$c+ immediately,
spawning a concurrent process that will calculate the \texttt{n}th Fibonacci
number and send it (denoted by the session type \lstinline+<!int;>+) along
\lstinline+$c+. Spawning functions provide a session type and a name preceded by
\lstinline+$+ for the returned channel so that the function body can use it
inside the body.
The spawned process is referred to
as the \emph{provider}, and the caller of the spawning function is referred to
as the \emph{client}.

\autoref{fig:fib.c1} provides an example with very simple session types to
demonstrate CC0's concurrent programming mechanisms. The client, \lstinline+main()+,
spawns a \lstinline+fib(10)+ provider and then receives on the
resulting channel. Note that the spawn \emph{does not} block the
\lstinline+main()+ process, but the receive does. Communication in CC0 is
asynchronous (channels are always buffered), so sends are always non-blocking,
but receives have to block until a message is available.

In the non-trivial case, \lstinline+fib(n)+ spawns two concurrent processes to
calculate \lstinline+fib(n-1)+ and \lstinline+fib(n-2)+. After receiving a
value, the parent \lstinline+fib(n)+ waits for the children to close their
channels (\lstinline+$c1+ and \lstinline+$c2+) before sending the result back
and then closing its own channel \lstinline+$c+. The compiler statically
verifies that sends and receives are performed in the proper order with the
proper types according the channel's session type (\autoref{sec:session-types}).
Also, the compiler makes sure that all channels are closed and properly waited
for (\autoref{sec:linearity}).

\subsection{Session Types}
\label{sec:session-types}

In concurrent programming, communication between two processes is often supposed
to follow some sort of protocol. By adding a type discipline to the (untyped)
$\pi$-calculus, session typing presents a method of encoding the type of this
communication: sequences of types represent how the type changes as the
communication takes place \cite{Caires10concur, Honda93concur, Toninho15phd}.
Each type in the sequence is designated as sending or receiving, encoding the
direction of communication. This captures the temporal aspect of concurrent
communication in a way that conventional (monotyped) channels do not: the type
actually reflects processes' progress in communicating with one another.

In Concurrent C0, session types are represented as a semicolon-separated
sequence of types between angle brackets. Each type is preceded by either
\lstinline+!+ or \lstinline+?+ to denote that a message of the given type is
sent or received, respectively. For example, a type where the provider sends an
\lstinline+int+, then a \lstinline+bool+, then receives an \lstinline+int+ would
be written as: \lstinline+<!int; !bool; ?int;>+. The final semicolon is
required; it indicates the end of communication along that channel.

Note how, in \autoref{fig:fib.c1}, the provider (\lstinline+fib+) behaves
according to \lstinline+<!int;>+, but the client (\lstinline+main+) does the
opposite, receiving where the other sends. Concurrent C0 uses dyadic session
types to model the client/provider relationship. Because both ends of the
channel communicate using the same protocol, it suffices to just give one type;
we type the session from the provider's point of view. The client will then have
to obey the \emph{dual} of that type. Duality is an important notion in session
typing that captures the requirement that communication actions occur in pairs:
if a provider sends an \lstinline+int+, the client must receive an
\lstinline+int+.

Session typing systems provide \emph{session fidelity}, the property
guaranteeing that processes send and receive the correct data in the correct
order according to the session type of the channel. For more on session typing,
duality, and session fidelity, see \cite{Caires10concur, Honda93concur,
Pfenning15fossacs}.

Session types allow bidirectional communication, but only in one direction at a
time. Consider process $A$ providing to client $B$ with the type
\lstinline+<?int; ?int; !bool;>+. The direction of communication starts out
toward the provider: $A$ is receiving and $B$ is sending. When $A$ has received
both \lstinline+int+s, $A$ has received everything $B$ sent but has not sent the
\lstinline+bool+ yet, so we know the channel buffer must be empty. Also, we know
that $B$ has sent both \lstinline+int+s, so its next action will be to receive;
both $A$ and $B$ are at the at the same point in the session type.

When session types change direction like this, a \emph{synchronization point}
occurs: both processes must be in the same place in the session type and the
buffer must be empty, allowing the direction of communication to switch.
Synchronization points occur whenever a session type change directions; a formal
treatment can be found in \cite{Pfenning15fossacs}.

\subsubsection{Branching}
\label{sec:branching}

Many protocols are not characterized by a straightforward sequence of types. CC0
uses the keyword \lstinline+choice+ to denote session types that branch into
different sequences of types. Choices are declared in a manner similar to
\lstinline+struct+s: a list of labels preceded by types. Using these constructs,
the C-like syntax can concisely express even complex session types.

Branches are selected by sending and receiving \emph{labels}, the values of
\lstinline+choice+ types. Labels are sent using dot notation:
\lstinline+$c.Label+. The \lstinline+switch+ operator is used to receive
values of \lstinline+choice+ types: when it takes a channel variable, it
receives and cases on the possible labels. The \lstinline+case+ branches must
follow the appropriate session type, as indicated by the label.

Receiving a label from the provider (\lstinline+?choice+) is called an
\emph{external choice}, because the client is making the decision. External
choices are a natural way to encode a server request: the client dictates the
type of action the server takes. The \lstinline+empty()+ function on
\autoref{line:queueEmpty} of \autoref{fig:queue.c1} offers an external choice.
Likewise, sending a label from the provider (\lstinline+!choice+) is an
\emph{internal choice}, because the provider specifies which branch the client
will take. Internal choices are ideal for encoding a server response, where the
client needs to react to different possibilities. The \lstinline+Deq+ branch on
\autoref{line:queueDeqType} in \autoref{fig:queue.c1} is an internal choice; the
client must handle the \lstinline+None+ case where no element is available.

In Concurrent C0, \lstinline+choice+s allow the user to name a session type,
also giving the ability to specify recursive ones. In \autoref{fig:queue.c1},
\lstinline+choice queue+ is a recursive session type for a provider that offers
a queue of integers. Once an element is enqueued, the type dictates that the
provider will continue to behave like a queue. Recursive session types can be
implemented with tail recursion (\autoref{fig:queue.c1},
\autoref{line:queueEmptyEnqElemTailCall}) or with loops. CC0 guarantees that a
new process is not spawned by a tail recursive call; it is executed in place by
the current provider.

\subsection{Linearity}
\label{sec:linearity}

Channel variables have linear semantics \cite{Caires10concur}, but with two
references: exactly one client and one provider will have a
reference to a channel. This ensures that communication is always one-to-one;
there can never be a ``dangling'' channel with no one listening on the other
end, nor will there ever be multiple providers or clients fighting to
communicate in one direction over a channel\footnotemark{}. Because a provider
can only have one client at a time (initially the caller of the function that
spawned it), there is a natural correspondence between the client-provider
relationship in a CC0 program and the parent-child relationship in a tree. The
\lstinline+main()+ function is a process with no clients and therefore the root
of the tree.

\footnotetext{
  CC0 implements \emph{linear channels} from \cite{Caires10concur,
  Pfenning15fossacs} which have exactly one client. The same paper provides a
  notion of \emph{shared channels} which can support multiple clients, but these
  are not presently in CC0.
}

The \lstinline+close+ and \lstinline+wait+ primitives let the provider and
client satisfy the linear type system. A process providing across channel
\lstinline+$c+ must call \lstinline+close($c)+ to terminate. The
provider must have already consumed all of its references and be a leaf in the
process tree. Before terminating, a client with a reference to a channel
\lstinline+$c+ must call \lstinline+wait($c)+ to ensure the provider terminates.

Channel references can be manipulated like other variables, but they are still
subject to linearity throughout the whole program. They cannot be copied, only
renamed; the old reference cannot be used. When passing a channel into a
spawning function, the caller gives up its reference to allow the new process to
use the channel. Sending channels along channels works much in the same way: the
sender gives up its reference to the receiving process. Linearity ensures that
channel references are not leaked or duplicated, so the process tree will remain
a tree even with dramatic manipulation of the communication structure.

\subsection{Forwarding}
\label{sec:forwarding}

Concurrent C0 implements an operation not commonly found in other languages with
message passing called \emph{forwarding} which allows a process to terminate
before its child and remove itself from the process tree. A node with exactly
one child\footnote{ Because forwarding terminates the process, linearity
dictates that all of its other references must have been properly destroyed at
the time of the forward.} can be contracted by the forward operation, allowing
its parent and child to communicate directly without it in the middle. Removing
the inner process effectively merges the two channels; because a process can
only forward channels of the same session type, session fidelity is preserved
and communication continues as if nothing happened \cite{Pfenning15fossacs,
Toninho15phd}.

\begin{wrapfigure}{R}{0.35\textwidth}
\vspace{-0.8cm}
\centering \footnotesize
  \begin{tikzpicture}[x=1.2cm, y=0.8cm]
    \node (P) at (0,0){$P$};
    \node (Q) at (1,0) {$Q$};
    \node (R) at (2,0) {$R$};
    \draw[o-] (Q) to node[below] {\lstinline{$c}} (P);
    \draw[o-] (R) to node[below] {\lstinline{$d}} (Q);

    \node (P2) at (3,0) {$P$};
    \node (R2) at (4,0) {$R$};
    \draw[o-] (R2) to node[below] {\lstinline{$c}} (P2);

    \draw[->,dashed,bend left=20] (1.5,0.5) to (3.1,0.5);

  \end{tikzpicture}
\vspace{-1em}
\end{wrapfigure}

At a very high level, forwarding can be thought of as setting a channel equal to
another channel. To the right, process $Q$ executes the forward \lstinline+$c = $d+,
terminating and combining the two channels into
one.
It is not obvious how to merge buffers that contain messages;
what if \lstinline+$d+ was not empty at the time of the forward?
See
\autoref{fig:forward-detail-dir-mismatch} for an example where
concatenation does not work, because messages are temporarily flowing in
opposite directions.


We propose an alternate view of forwarding: as a special kind of message. We use
the session typing system to infer the direction of communication according to
the forwarding process, so a forward sends a special message along the channel
in that direction containing a reference to the other channel. This message must
be the last one in the buffer because the forwarding process terminated after
sending it. When a process receives the forward, it destroys the channel it was
sent over and replaces its reference with the new channel from the forward
message. \autoref{fig:forward-as-message} contains a more detailed example of
how forwarding works.


\begin{figure}

  \begin{minipage}[t][\textheight][b]{0.4\textwidth}
    \scriptsize
\begin{lstlisting}[
    xleftmargin=1cm,
    numbers=left,
    numberstyle=\tiny\color{gray},
    ]
// external choice for request
choice queue {
  <?int; ?choice queue>  Enq;
  <!choice queue_elem>   Deq; /*% \label{line:queueDeqType} %*/
};
typedef <?choice queue> queue;

// internal choice for response
choice queue_elem {
  <!int; ?choice queue> Some;
  < >                   None;
};

// provider that holds element x and
// points to the rest of the queue $r
queue $q elem (int x, queue $r) {
  switch ($q) {
  case Enq:
    int y = recv($q);
    $r.Enq; send($r, y);  /*% \label{line:queueElemEnqPass} %*/
    $q = elem(x, $r);
  case Deq:               /*% \label{line:queueElemDeq} %*/
    $q.Some; send($q, x); /*% \label{line:queueElemDeqRespond} %*/
    $q = $r;              /*% \label{line:queueElemDeqForward} %*/
  }
}

// provider for the end of the queue
queue $q empty () {     /*% \label{line:queueEmpty} %*/
  switch ($q) {
  case Enq:
    int y = recv($q);
    queue $e = empty(); /*% \label{line:queueEmptyEnqEmptySpawn} %*/
    $q = elem(y, $e);   /*% \label{line:queueEmptyEnqElemTailCall} %*/
  case Deq:
    $q.None;
    close($q);
  }
}

void dealloc (queue $q) {
  $q.Deq; switch($q) { /*% \label{line:queueMainDeq} %*/
    case Some:
      recv($q);
      dealloc($q);
      return;
    case None:
      wait($q);
      return;
  }
}

int main () {
  queue $q = empty();
  $q.Enq; send($q, 1); /*% \label{line:queueMainEnq1} %*/
  $q.Enq; send($q, 2); /*% \label{line:queueMainEnq2} %*/
  dealloc($q);
  return 0;
}
\end{lstlisting}
    \vfill
    \caption{\texttt{queue.c1}, a queue implementation where each element is a
      concurrent process.}
    \label{fig:queue.c1}
  \end{minipage}%
  \hfill
  \begin{minipage}[t][\textheight][b]{0.55\textwidth}
    \footnotesize
    \caption*{\tiny
      The arrows above the channel contents indicate the actual flow of
      messages along the channel. The small arrows above channel endpoints
      indicate the direction of that process' next action along that channel
      according the session type.}
    \vspace{\fill}

    \begin{subfigure}{\linewidth}
    \centering
      \begin{tikzpicture}
        \node (P) {$P$};
        \node (Pproc) [below=0cm of P] {\lstinline+main()+};
        \node (Q) [right=2.5cm of P] {$Q$};
        \node (Qproc) [below=0cm of Q] {\lstinline+empty()+};
        \draw[-o] (P) to
          node[below] {\lstinline+$q+}
          node[above,at start] {$\rightarrow$}
          node[above] {\(
            \overrightarrow{
              \begin{tabular}{|c|c|}
                \hline 1 & \lstinline+Enq+ \\ \hline
              \end{tabular}
            }\)}
          node[above,at end] {$\rightarrow$}
          (Q);
      \end{tikzpicture}
      \caption{$P$ enqueues 1 [\autoref{line:queueMainEnq1}].}
    \end{subfigure}
    \vspace{\fill}

    \begin{subfigure}{\linewidth}
    \centering
      \begin{tikzpicture}
        \node (P) {$P$};
        \node (Pproc) [below=0cm of P] {\lstinline+main()+};
        \node (Q) [right=2.5cm of P] {$Q$};
        \node (Qproc) [below=0cm of Q] {\lstinline+elem(1,$r)+};
        \node (R) [right=2.5cm of Q] {$R$};
        \node (Rproc) [below=0cm of R] {\lstinline+empty()+};
        \draw[-o] (P) to
          node[below] {\lstinline+$q+}
          node[above,at start] {$\rightarrow$}
          node[above,at end] {$\rightarrow$}
          (Q);
        \draw[-o] (Q) to
          node[below] {\lstinline+$r+}
          node[above,at start] {$\rightarrow$}
          node[above,at end] {$\rightarrow$}
          (R);
      \end{tikzpicture}
      \caption{$Q$ gets the enqueue, spawning a new \lstinline+empty()+
        process and channel [\autoref{line:queueEmptyEnqEmptySpawn}--%
        \ref{line:queueEmptyEnqElemTailCall}].}
    \end{subfigure}
    \vspace{\fill}

    \begin{subfigure}{\linewidth}
    \centering
      \begin{tikzpicture}
        \node (P) {$P$};
        \node (Pproc) [below=0cm of P] {\lstinline+main()+};
        \node (Q) [right=2.5cm of P] {$Q$};
        \node (Qproc) [below=0cm of Q] {\lstinline+elem(1,$r)+};
        \node (R) [right=2.5cm of Q] {$R$};
        \node (Rproc) [below=0cm of R] {\lstinline+empty()+};
        \draw[-o] (P) to
          node[below] {\lstinline+$q+}
          node[above,at start] {$\leftarrow$}
          node[above] {\(
            \overrightarrow{
              \begin{tabular}{|c|}
                \hline \lstinline+Deq+ \\ \hline
              \end{tabular}
            }\)}
          node[above,at end] {$\rightarrow$}
          (Q);
        \draw[-o] (Q) to
          node[below] {\lstinline+$r+}
          node[above,at start] {$\rightarrow$}
          node[above] {\(
            \overrightarrow{
              \begin{tabular}{|c|c|}
                \hline 2 & \lstinline+Enq+ \\ \hline
              \end{tabular}
            }\)}
          node[above,at end] {$\rightarrow$}
          (R);
      \end{tikzpicture}
      \caption{$P$ enqueues 2 [\autoref{line:queueMainEnq2}] which $Q$ passes to the
        back of the queue [\autoref{line:queueElemEnqPass}]. $P$ sends a dequeue
        request and waits for the result [\autoref{line:queueMainDeq}].}
    \end{subfigure}
    \vspace{\fill}

    \begin{subfigure}{\linewidth}
    \centering
      \begin{tikzpicture}
        \node (P) {$P$};
        \node (Pproc) [below=0cm of P] {\lstinline+main()+};
        \node (Q) [right=2.5cm of P] {$Q$};
        \node (R) [right=2.5cm of Q] {$R$};
        \node (Rproc) [below=0cm of R] {\lstinline+empty()+};
        \draw[-o] (P) to
        node[below] {\lstinline+$q+}
        node[above,at start] {$\leftarrow$}
        node[above] {\(
          \overleftarrow{
            \begin{tabular}{|c|c|}
              \hline \lstinline+Some+ & 1 \\ \hline
            \end{tabular}
          }\)}
        node[above,at end] {$\rightarrow$}
        (Q);
        \draw[-o] (Q) to
        node[below] {\lstinline+$r+}
        node[above,at start] {$\rightarrow$}
        node[above] {\(
          \overrightarrow{
            \begin{tabular}{|c|c|}
              \hline 2 & \lstinline+Enq+ \\ \hline
            \end{tabular}
          }\)}
        node[above,at end] {$\rightarrow$}
        (R);
      \end{tikzpicture}
      \caption{$Q$ responds to the dequeue [\autoref{line:queueElemDeqRespond}] and
        is about to forward [\autoref{line:queueElemDeqForward}].}
    \end{subfigure}
    \vspace{\fill}

    \begin{subfigure}{\linewidth}
    \centering
      \begin{tikzpicture}
        \node (P) {$P$};
        \node (Pproc) [below=0cm of P] {\lstinline+main()+};
        \node (Q) [right=2.5cm of P] {$\xcancel{Q}$};
        \node (R) [right=3.2cm of Q] {$R$};
        \node (Rproc) [below=0cm of R] {\lstinline+empty()+};
        \draw[-o] (P) to
        node[below] {\lstinline+$q+}
        node[above,at start] {$\leftarrow$}
        node[above] {\(
          \overleftarrow{
            \begin{tabular}{|c|c|}
              \hline \lstinline+Some+ & 1 \\ \hline
            \end{tabular}
          }\)}
        (Q);
        \draw[-o] (Q) to
        node[below] {\lstinline+$r+}
        node[above] {\(
          \overrightarrow{
            \begin{tabular}{|c|c|c|}
              \hline \lstinline+fwd: $q+ & 2 & \lstinline+Enq+ \\ \hline
            \end{tabular}
          }\)}
        node[above,at end] {$\rightarrow$}
        (R);
      \end{tikzpicture}
      \caption{$Q$ forwards \lstinline+$q = $r+
        [\autoref{line:queueElemDeqForward}] by sending a forward message in the
        direction of communication according to the session type, not the state
        of the channel buffers. $Q$ terminates, but \lstinline+$r+ must persist
        because it still has messages. Simply concatenating the buffer here will
        not work because they have different directions.}
      \label{fig:forward-detail-dir-mismatch}
    \end{subfigure}
    \vspace{\fill}

    \begin{subfigure}{\linewidth}
    \centering
      \begin{tikzpicture}
        \node (P) {$P$};
        \node (Pproc) [below=0cm of P] {\lstinline+main()+};
        \node (R) [right=2.5cm of P] {$R$};
        \node (Rproc) [below=0cm of R] {\lstinline+elem(2,$s)+};
        \node (S) [right=2.5cm of R] {$S$};
        \node (Sproc) [below=0cm of S] {\lstinline+empty()+};
        \draw[-o] (P) to
          node[below] {\lstinline+$q+}
          node[above,at start] {$\rightarrow$}
          node[above,at end] {$\rightarrow$}
          (R);
        \draw[-o] (R) to
          node[below] {\lstinline+$s+}
          node[above,at start] {$\rightarrow$}
          node[above,at end] {$\rightarrow$}
          (S);
      \end{tikzpicture}
      \caption{$R$ finally gets the enqueue, spawning a new \lstinline+empty()+
        process and channel [\autoref{line:queueEmptyEnqEmptySpawn}--%
        \ref{line:queueEmptyEnqElemTailCall}]. When $R$ receives the forward, it
        deallocates \lstinline+$r+ (which is now safe because \lstinline+$r+ is
        empty) and will now use \lstinline+$q+ instead.}
    \end{subfigure}
    \vspace{\fill}

    \caption{An illustration using \texttt{queue.c1} (\autoref{fig:queue.c1})
      demonstrating how treating forwarding as a message resolves communication
      direction issues.}
    \label{fig:forward-as-message}
  \end{minipage}

\end{figure}


\autoref{sec:forwarding-implementation} discusses the details of our
implementation, but it's important to note that this interpretation of
forwarding allows implementation on any level. This view of forwarding, to the
best of our knowledge, is a novel contribution of this work, and could be
implemented in any session typed, message passing language.


\section{Implementation}
\label{sec:impl}

Concurrent C0's typing system not only ensures the safety of concurrent code,
but it also allows for an efficient parallelizable implementation. Session
typing directly enables our implementation to use fewer, smaller buffers than
other message passing techniques.

\subsection{Compiler}

Concurrent C0 enforces linearity and session fidelity to produce safe concurrent
code. The compiler typechecks programs to make sure that messages are sent and
received according to the appropriate session types, and it also ensures that
the linearity of channels is respected. While certainly important to CC0, the
typechecker itself is not a novel contribution of this work, and interested
readers are referred to \cite{Griffith16phd} for more about
typechecking session typed and linear languages. After typechecking, the
compiler inserts annotations that inform the runtime about the communication
structure. Finally, CC0 source code is compiled to a target language (C or Go)
then linked with a runtime implementation written in the same target language.

\subsubsection{Type Width}
\label{sec:type-width}

Certain session types dictate that only so many values can be buffered at a
time. For example, the type \lstinline+<!bool; ?int;>+ could only possibly
buffer one value at a time, because the \lstinline+int+ must be sent from
the client, which can only occur once the client has received the previous
\lstinline+bool+. This quantity is called the \emph{width} of the type. The CC0
compiler infers widths, allowing the runtime to use small, fixed length circular
buffer as queues and not have to worry about ever resizing.

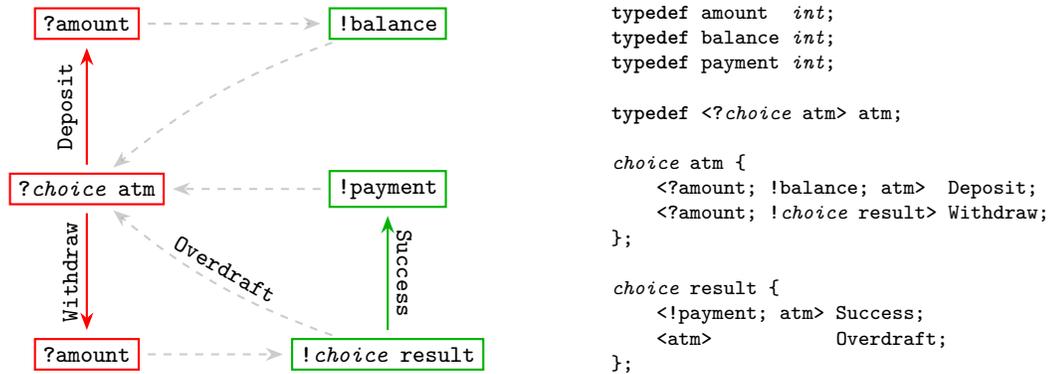
\begin{figure}
  \centering
  \begin{subfigure}{0.59\textwidth}
  \centering
    \begin{tikzpicture}[
        x=4cm, y=2.2cm,
        thick,
        -{Stealth[scale=1.0]}, shorten >=3pt, shorten <=3pt,
        send/.style={draw=black!30!green},
        recv/.style={draw=red},
        sync/.style={draw=gray, dashed, opacity=0.4}]

      \footnotesize

      \node[recv] (atm)             at (0, 0) {\lstinline+?choice atm+};
      \node[recv] (Deposit-amount)  at (0, 1) {\lstinline+?amount+};
      \node[recv] (Withdraw-amount) at (0,-1) {\lstinline+?amount+};
      \node[send] (result)          at (1,-1) {\lstinline+!choice result+};
      \node[send] (Success-payment) at (1, 0) {\lstinline+!payment+};
      \node[send] (Deposit-balance) at (1, 1) {\lstinline+!balance+};

      \path[anchor=south]
        (atm)             edge[recv] node[rotate=90]  {\lstinline+Deposit+}  (Deposit-amount)
                          edge[recv] node[rotate=90]  {\lstinline+Withdraw+} (Withdraw-amount)
        (Deposit-amount)  edge[sync]                                          (Deposit-balance)
        (Deposit-balance) edge[sync, bend right=10]                           (atm)
        (Withdraw-amount) edge[sync]                                          (result)
        (Success-payment) edge[sync]                                          (atm)
        (result)          edge[send] node[rotate=270] {\lstinline+Success+}  (Success-payment)
                          edge[sync, bend left=10]
                          node[sloped, draw=none, opacity=1] {\lstinline+Overdraft+} (atm);
    \end{tikzpicture}
    \label{fig:atm-graph}
  \end{subfigure}\begin{subfigure}{0.4\textwidth}
    \scriptsize
\begin{lstlisting}
typedef amount  int;
typedef balance int;
typedef payment int;

typedef <?choice atm> atm;

choice atm {
    <?amount; !balance; atm>  Deposit;
    <?amount; !choice result> Withdraw;
};

choice result {
    <!payment; atm> Success;
    <atm>           Overdraft;
};
\end{lstlisting}
    \label{fig:atm-code}
  \end{subfigure}
  \caption{Code and graph of type \lstinline+atm+ with width 2.}
  \label{fig:atm}
\end{figure}

Session types can be viewed as a directed graph in which a walk represents a
possible sequence of sent or received types. Nodes are colored as sending
(green) or receiving (red); see \autoref{fig:atm}. We know that the buffer will
only contain messages going in one way at a time, so there are actually two
graphs, one red and one green, connected by the dashed gray edges representing
synchronization points where we know the buffer will be empty. Thus, the width
of the type is the number of nodes in the longest walk in either subgraph.

An ATM is a canonical example in the session typing literature, and
\autoref{fig:atm} shows the code and type graph for a simple ATM
protocol. A process providing \lstinline+<?choice atm>+ could clearly stay alive
forever: the client may \lstinline+Deposit+ or \lstinline+Withdraw+ an unbounded
number of times. However, the width of the type is only 2; so the channel
will never need to buffer more than two items.

Note that type width is compatible with forwarding because of the
forward-as-message interpretation. Forwards are received \emph{instead of} the
intended message, so the forward message will occupy that allocated space.

\subsubsection{Forwarding}
\label{sec:forwarding-implementation}

At each forward call-site, the CC0 compiler infers the direction of
communication according the forwarding process's session type. In the generated
code, that direction is passed into the forward runtime function. Just like the
semantic understanding, the runtime sends a specially tagged message in that
direction and then terminates the calling process. Nothing else occurs until the
forward is received.

A process attempting to receive another value may see the special forward tag
instead. The forward message is guaranteed to be the last message in the buffer,
so the receiving process destroys the channel. The forward message contains a
reference to the new channel, so the receiving process replaces its own
reference to the destroyed channel with the new one, and then it attempts to
receive the value it initially expected over that new channel. This ensures the
transparency of the forward: this process is still going to receive the value
that it expected, and all future interactions over that channel reference will
use the new channel instead. Because forwards are deferred, the receiver may 
need to handle many forwards before getting the expected message.

\subsection{Runtime System}

The CC0 compiler generates C or Go code that is linked with one of several
runtime systems that contain the logic for message passing and manipulating
processes. The runtimes have different threading models and synchronization
strategies, but they share the same general structure centered around channels.
A channel contains a message queue, its direction, a mutex, and a condition
variable. The mutex is necessary to protect channel state, and the condition
variable is used by receivers to wait on messages to arrive or the queue to
change directions.

The runtime consists of four main functions that provide all the necessary
functionality for spawning processes and message passing:
\lstinline+NewChannel+, \lstinline+Send+, \lstinline+Recv+, and
\lstinline+Forward+. The CC0 functions \lstinline+close+ and \lstinline+wait+ are
implemented by sending and receiving a special \texttt{DONE} message.

\lstinline+NewChannel+ creates a new concurrent process and the channel along
which it will provide, returning a reference to that channel to the caller
(client). \lstinline+NewChannel+ takes in the function and arguments for the new
provider process as well as the type width and initial direction of the channel
as inferred by the compiler, allowing the runtime to create a channel with a
bounded ring-buffer when possible.

\lstinline+Send+ sends a given message over a given channel, additionally taking
in the message's type and the inferred direction. \lstinline+Send+ locks the
channel, enqueues the message with its type, sets the direction of the queue,
and unlocks. A receiver may be waiting for the message, so the sender must wake
up the potential receiver by signaling the condition variable.

\lstinline+Recv+ receives a message over a given channel, taking in the
message's expected type and the inferred direction. \lstinline+Recv+ locks the
channel and attempts to receive the message. If the buffer is empty or still
flowing in the other direction, then the caller will give up the lock and wait
on the condition variable for the sender. If the message is a forward, the
receiver handles it, installing the new channel; see
\autoref{sec:forwarding-implementation} for details. \lstinline+Recv+ asserts
that the received message is of the expected type, panicking if it does not
match (and is not a forward).

\lstinline+Forward+ takes in the two channels involved in the forward and the
inferred direction of communication. The forwarding process sends a forward
message in the inferred direction using the regular message passing
functionality, then it terminates. See \autoref{sec:forwarding} and
\autoref{sec:forwarding-implementation}.


Because CC0 encourages highly concurrent programming, most programs spawn many
processes. Our early C runtimes used 1:1 or custom M:N threading models; neither
performed as well as the Go runtimes. Go is an imperative programming language with
a lightweight threading model that provides concurrency and efficient
parallelism \cite{Go}.



\section{Experimental Comparison and Analysis}
\label{sec:analysis}

To benchmark Concurrent C0, we created \texttt{go0}, a naive, proof-of-concept
implementation that uses Go's built-in channels to implement CC0 channels. As
CC0 channels provide safe bidirectional communication, two Go channels must be
used to implement a CC0 channel without additional synchronization. \texttt{go0}
serves as a stand-in modeling how message passing is done in other languages
(with two large channels intended for one-way communication), but it conforms to
the same interface as our other implementations so we can run the same tests
against it.

Analysis of both the C and Go runtimes can be found in the
\href{http://maxwillsey.com/papers/cc0-thesis.pdf}{extended version} of
this paper; here we compare \texttt{go0} against \texttt{go2}, a Go
implementation which uses the full suite of language based optimizations
detailed in \autoref{sec:impl}. Our benchmarking suite\footnotemark{} consists
of many highly concurrent data structures, like the queue in
\autoref{fig:queue.c1}. Most of the work done in these tests is communication,
so as to highlight the efficiency of our message passing runtimes. All
benchmarks were run on a 2015 MacBook Pro with an Intel Core i7-4870HQ CPU with
4 cores at 2.50GHz.
\footnotetext{ See the full suite of tests at
\url{http://maxwillsey.com/assets/cc0-linear16-benchmarks.tgz}}

\begin{figure*}
  \centering
  \includegraphics[width=1.0\textwidth]{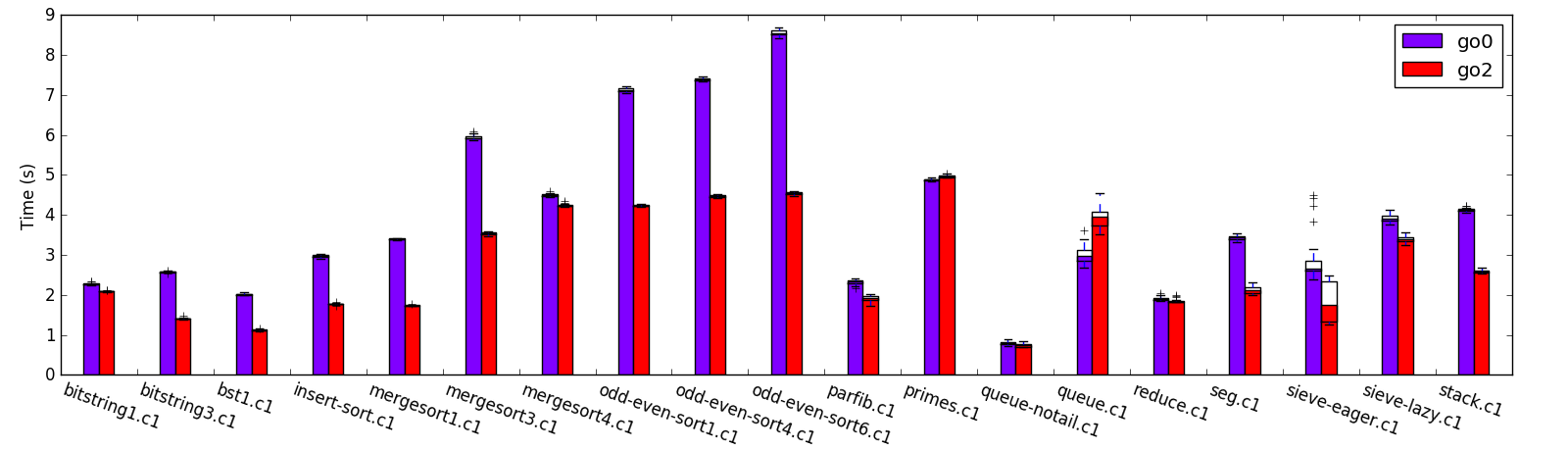}
  \caption{Median benchmark times of the \texttt{go0} and \texttt{go2} runtimes
    over 20 samples.}
  \label{fig:goBenchmarks}
\end{figure*}

The \texttt{go0} vs. \texttt{go2} benchmark in \autoref{fig:goBenchmarks}
demonstrates the effectiveness of our implementation techniques. Compared to the
naive implementation, our optimized version ran $1.38\times$ faster on
average\footnote{
Average is calculated as geometric mean of the ratios of the median benchmark
times}.
We suspect that the speed up would be even more dramatic if the Go compiler
optimized tail recursion. Concurrent C0 encourages a tail recursive style of
programming; see \autoref{sec:branching}. The \texttt{queue-notail.c1} test case
is the same as \texttt{queue.c1}, except that it is written with loops as
opposed to tail recursion. The more than $2\times$ difference in both Go
runtimes' performance indicates that Go's lack of optimization in this case is a
serious hindrance. Other test cases like \texttt{primes.c1} rely heavily on
mutually recursive tail calls, so even though the negative impact is similar, no
\texttt{-notail} version was written for those cases.


\section{Future Work}
\label{sec:conclusion}
\label{sec:distributed}  
\label{sec:sharedmemory} 

The knowledge given by session types could improve scheduling decisions. The
structure of relationships between communicating process could enable
optimizations like co-scheduling providers and clients to increase parallel
performance. The same information might also assist the runtime in deciding on
granularity; the structure of the process tree could help save the overhead of
spawning new concurrent processes in some situations, just running them inline
instead.

Session typing and linearity make communication of values safe, and Concurrent
C0 (from C0) is memory safe for sequential programs, but the combination of
shared memory and concurrency leads to race conditions. Presently, our
implementations allow sending and receiving pointer and arrays between
processes, but there is no attempt to enforce the safety of accesses and writes.
Given that channels already have linear semantics, CC0 could benefit from a
linear or affine treatment of shared memory like that of Rust\footnote{
\url{https://doc.rust-lang.org/book/ownership.html}}.

Session types are traditionally associated with distributed computing, so a
distributed implementation of Concurrent C0 could apply some of the
contributions of this work. Specifically, the concept of forwarding as a message
would be even more beneficial than it was in the shared memory setting, as
synchronization is even more challenging on the distributed scale.






\newpage
\defbibheading{noheader}[\refname]{\section*{#1}}                                                   
\printbibliography[heading=noheader] 

\end{document}